\documentclass{article}    

\usepackage{graphicx}

\title{\textbf{Optical Shelving: Suppressed Fluorescence}}  
\author{Richard A. Mould\footnote{Department of Physics and Astronomy, State University of New York, Stony Brook,
\mbox{New York} 11794-3800; http://ms.cc.sunysb.edu/\~{}rmould}}  
\date{}    
   
\begin{document}             

\maketitle              

\begin{abstract}

The shelving phenomenon of quantum optics, originally observed by Dehmelt, is analyzed in terms of the qRules that are given in another paper.  The heuristic value of these rules is apparent because they not only describe the dark period during shelving, but they reveal the mechanism that enforces the suppression of fluorescence during that time.

\end{abstract}

\section*{Introduction}
 
Given an atom with three energy levels $a_0, a_1$, and $a_2$, where $a_0$ is the ground state and $a_1$ and $a_2$ are excited
states.  The atom is exposed to two laser beams, one of excitation energy 0-1 and the other of excitation energy 0-2, where
$a_2$ is a much longer lived state than $a_1$; so the 0-1 photons are stronger than the  0-2 photons.  At time $t_0$ the atom
begins in its ground state.

The atom will respond with the release of a strong photon.  It then resets to ground and repeats the process, emitting another
strong photon.  This continues for a time called the \emph{fluorescent period} during which a shower of many strong photons are
rapidly released.  Weak 0-2 photons
do not appear during the fluorescent period.  However, after a time the weak interaction does prevail, blocking the fluorescence
and initiating a \emph{dark period} that lasts for  the half-life of a weak photon.  Dehmelt originally
explained this by saying that the atom occasionally jumps to the $a_2$ state where it is \emph{shelved} until it decays again to
ground.  The atom is then \emph{fully reset} to ground emitting a  photon, and a fluorescent period begins again followed in time by
another dark period \cite{HD, PLK, PP}.  

It  is not immediately clear how
the weak interaction manages to cut off all fluorescent photons for so long a period of time.  Why doesn't  fluorescence
always override the possibility of an occasional weak photon?  This is the question raised by Shimony and others \cite{AS}.  It is the
purpose of this paper to answer this question using the qRules that are claimed by the author to be auxiliary to Schr\"{o}dinger's equation.  These rules are interpretive of quantum mechanical equations and processes \cite{RM}.  

The Schr\"{o}dinger solution to the shelving problem is given  by T. Erber   et al. \cite{TE} and is of the form
\begin{eqnarray}
a_0(t) &=&   cos[\Omega t] exp[-\beta t] + A\hspace{0.1cm} exp[i\Omega t] \{exp[-\beta t] - exp[-\lambda t]\}\\
a_1(t) &=& i \hspace{0.1cm} sin[\Omega t] exp[-\beta t] + A\hspace{0.1cm} exp[i\Omega t] \{exp[-\beta t] - exp[-\lambda t]\}\\
a_2(t) &=& \hspace{2.7cm}  -iB \hspace{0.1cm} exp[i\Omega t] \{exp[-\beta t] - exp[-\lambda t]\} 
\end{eqnarray}
where $\Omega$ is the  frequency and $\beta$ is the  decay constant of the strong interaction that produces fluorescence.  The cosine in Eq.\ 1 and
the sine in Eq.\ 2 identify this oscillation.  There is no similar `two-state' oscillation
involving the 0-2 transition.  Instead, there is a \emph{three-state resonance} given by exponential  
\begin{displaymath}
exp[i\Omega t]\{exp[-\beta t] - exp[-\lambda t]\}
\end{displaymath}
 This gives rise to the ``dark period" where the slow decay constant $\lambda$ insures a long
half-life.  When the sin/cos fluorescent components are extinguished, the remaining three-state resonance persists without (an immediate)
radiation decay.  Equations 1-3 do not include the reset radiation  components so they do not preserve normalization over time.
However, ShimonyÕs question is still not answered.   That is: What is the mechanism that suppresses fluorescence during the dark period?  If the atom is not `shelved' during this time as claimed by Dehmelt, then what enforces the fluorescent cut-off?

\section*{A qRule Analysis}
	The qRules are three  rules that govern the behavior of quantum mechanical systems beyond the dynamic prnciple.  They are listed in Ref. 5 together with examples involving microscopic systems as well as macroscopic systems, with or without an observer.  They are asssumed to be universal and   are applied below to the shelving problem. A photon detector is not present because we assume that the shelving phenomena described above is objective -- it does not depend on an external detector or observer making a so-called ``null measurement". The initial state of the system at $t_0$ is given by 
\begin{equation}
\Phi(t_0) = \gamma_n \gamma'_ma_0
\end{equation}
The radiation field contains
$n$ strong photons $\gamma_n$ with a frequency between the levels 0 and 1, and $m$ weak photons $\gamma'_m$ of
 frequency between the levels 0 and 2.  

  After $t_0$  Eqs. 1-3 are represented by the qRule equation
\begin{eqnarray}
\Phi(t \ge t_0)= \\ 
 &\leftrightarrow&\gamma_{n-1}\gamma'_m a_1 + \gamma_{n-1}\gamma'_m\underline{a}_0\otimes \gamma \hspace{1cm}\textrm{fluorescence} \hspace{2cm}\nonumber
\\ 
=\gamma_n\gamma'_ma_0&\leftrightarrow&\nonumber
\\
&\leftrightarrow&  \gamma_{n-1}\gamma'_ma_1 \leftrightarrow \gamma_n\gamma'_{m-1}a_2 + \gamma_{n-1}\gamma'_m\underline{a}_0\otimes\gamma +
\gamma_n\gamma'_{m-1}\underline{a}_0
\otimes\gamma'\nonumber
\end{eqnarray}
\hspace{3.3cm}three state resonance\hspace{1.8cm} full-reset states

\vspace{.29cm}
The initial component $\gamma_n\gamma'_ma_0$ oscillates (double arrows) with both the top  (fluorescence) row making a two-state resonance and  the bottom (full-reset) row making a  three-state resonance.   The components in these two rows are equal to zero at $t_0$.  The part of $\gamma_n\gamma'_ma_0$ that oscillates with the three-state resonance has the same amplitude as the first component in the bottom row, so it too is zero at $t_0$.  The laser induced two-state oscillation between $a_0$ and $a_1$ is given by the sin/cos components in \mbox{Eqs.\ 1-3}.  

The last component in the top row of Eq.\ 5 represents the spontaneous emission (indicated by $\otimes$) of a  photon $\gamma$ and a return of
the atom to ground.  It is called a \emph{ready component} as indicated by the underline of one of its states (in this case
$\underline{a}_0$).  Only a ready component is a candidate for state reduction according to the qRules.  With  probability
current flowing into it, a ready component is subject to a stochastic hit at each moment of time with a probability equal to the
current times $dt$.  All components except the chosen one are then reduced to zero.    After being chosen in this way a ready component is  no longer `ready'  (it is now called \emph{realized component}) and is no longer underlined.  

If the ready component in the top row is stochastically chosen at some \mbox{time $t_{sc}$},  a wave collapse will yield a new solution
 given by
\begin{equation}
\Phi(t = t_{sc} \ge t_0) = \gamma_{n-1}\gamma'_ma_0
\otimes\gamma\end{equation}
which is the same as Eq.\ 4 except that one of the $\gamma$ photons has been removed from the laser beam and has become a radiated
florescent photon.  The top row in \mbox{Eq.\ 5} is repeated many times during the fluorescent period.  

The bottom row of Eq.\ 5 contains the radiationless three-state resonance. Probability current is `stored' there until it is released through a spontaneous decay to one of the two full-reset states at the end of the that row.

Although the initial state $\gamma_n\gamma'_ma_0$ contributes to both resonances, current flows much faster into the top row than it does into the bottom row.  Also, the top row decays more rapidly.  This means that the system is more likely to fluorescent decay before it has a chance to fully reset.
	
Sooner or later the initial state contribution to the two-state resonance will become depleted before it can decay, inasmuch as   part of the initial state is tied up in the three-state resonance.  At that point Eq. 5 becomes
\begin{eqnarray}
\Phi(t \ge t_0)= 
\\ 
 &\leftrightarrow&0 + \gamma_{n-1}\gamma'_m\underline{a}_0\otimes \gamma \hspace{1cm}\textrm{no 	fluorescence} \hspace{2cm}\nonumber
\\ 
=\gamma_n\gamma'_ma_0&\leftrightarrow&\gamma_{n-1}\gamma'_ma_1 \leftrightarrow \gamma_n\gamma'_{m-1}a_2 + \gamma_{n-1}\gamma'_m\underline{a}_0\otimes\gamma +
\gamma_n\gamma'_{m-1}\underline{a}_0
\otimes\gamma'\nonumber
\end{eqnarray}
\hspace{2.3cm}three state resonance\hspace{2.8cm} full-reset states

\vspace{.3cm}
\noindent
where the remaining ready component in the top row is dormant and serves no further purpose.  At this point the bottom row will resonate at leisure, resulting in the dark period during which there is no fluorescence.  That resonance  decays with a long half-life $\lambda$ and finally discharges through a spontaneous decay to one of the  full-reset states at the end of  that row.  To the extent that $a_1$ in the three-state resonance is not zero it will decay  to the first      full-reset component; and to the extent that $a_2$ in that resonance is not zero it will decay to the second full-reset component.  With the stochastic choice of one of these two ready components there will be a full reset that completes the dark period with the emission of a $\gamma$ or a $\gamma'$ photon.

	This answers ShimonyÕs question as to the mechanism that cuts off the fluorescent radiation during the long dark period.  Fluorescent radiation is cut off because the initial component no longer feeds the two-state resonance.  What remains of that component is engaged in the three-state resonance, and the only escape from that resonance to a full reset is through a long half-life spontaneous photon emission.

	It is to be emphasized that the shelving phenomena described here is an objective property of the system and is not in any way dependent on the presence of an external detector or observer.  The idea that the existence of a dark period depends `causally' on the failure of a detector to see fluorescence makes no sense.  A ``null measurement" does not \emph{produce} a dark period; rather, it is only a \emph{consequence} of a dark period.

\end{document}